\documentstyle[12pt,epsf,epsfig]{article}
\begin{document}

\title{X(1835): A Natural Candidate of $\eta^\prime$'s Second Radial Excitation}
\author{Tao Huang$^{1,2}$ and Shi-Lin Zhu$^3$, \\
$^1$CCAST (World Laboratory), P. O. Box 8730, Beijing 100080,
China\\
$^2$Institute of High Energy Physics, Chinese Academy of Science\\
P. O. Box 918(4), Beijing 100049, China\\
$^3$Department of Physics, Peking University, Beijing 100871,
China} \maketitle

\begin{abstract}

Recently BES collaboration observed one interesting resonance
X(1835). We point out that its mass, total width, production rate
and decay pattern favor its assignment as the second radial
excitation of $\eta^\prime$ meson very naturally.

\end{abstract}
\vspace{0.3cm}

\pagenumbering{arabic}

\section{Introduction}

A significant $p\bar p$ threshold enhancement was reported by BES
Collaboration in the radiative decay $J/\psi\to \gamma p\bar p$
\cite{bes}. No similar signal was observed in the channel $\pi^0
p\bar p$. Assuming this enhancement arose from a resonance below
threshold, the central value of the assumed resonance from S-wave
fit was around 1859 MeV \cite{bes}. This year BES Collaboration
observed a new resonance X(1835) in the $J/\psi \to \gamma
\eta^\prime \pi^+\pi^-$ channel with a statistical significance of
$7.7\sigma$ \cite{bes1}. The $\eta^\prime$ meson was detected in
both $\eta\pi\pi$ and $\gamma\rho$ channels. There are roughly
$264\pm 54$ events. Its mass is $m_X=(1833.7\pm 6.2\pm 2.7)$ MeV
and its width is $\Gamma \left(X(1835)\right)=(67.7\pm 20.3\pm
7.7)$ MeV \cite{bes1}.

There are many speculations of the underlying structure of the
$p\bar p$ threshold enhancement and X(1835) in literature.
Proposed theoretical schemes include the t-channel pion exchange,
some kind of threshold kinematical effects, a new resonance below
threshold, even a $p\bar p$ bound state etc
\cite{rosner,gao,q1,q2,q3,q4,q5,q6,q7,q8,q9,q10,q11,q12,q13,q14,zhu,q15}.

The possibility of the $p\bar p$ threshold enhancement being a
pseudoscalar glueball was discussed extensively in Ref.
\cite{koch}, and later in Refs. \cite{hxg,lba}. One serious
obstacle of this assignment is its low mass. Lattice QCD predicts
the pure scalar glueball around $1.5\sim 1.7$ GeV \cite{lattice}.
Experimentally there exist overpopulation of scalar mesons around
$1.3\sim 1.7$ GeV. Pure pseudoscalar glueballs are predicted to
lie around 2.6 GeV \cite{lattice}. Therefore one needs to find a
special powerful mixing mechanism to pull its mass from 2.6 GeV
down to 1.835 GeV.

Under the strong assumption that the $p\bar p$ threshold
enhancement and X(1835) are the same resonance, Zhu and Gao
suggested X(1835) could be a $J^{PC}I^G=0^{-+}0^+$ $p\bar p$
baryonium \cite{zhu}. Such a scheme easily explains the large
branching ratio of $X\to p\bar p$ observed by BES. Moreover, the
dominant decay modes are $X(1835)\to \eta \pi\pi$ and $X(1835)\to
\eta^\prime \pi\pi$. Three-body decay modes with strangeness are
suppressed due to the absence of explicit strangeness within a
$p\bar p$ baryonium \cite{gao,zhu}. BES collaboration did observe
X(1835) in the $\eta^\prime \pi\pi$ channel. However they have not
reported any positive information on the $\eta \pi\pi$ mode. The
latter mode should have bigger branching ratio if X(1835) is a
$p\bar p$ baryonium \cite{zhu}. If future experimental search
fails to observe X(1835) in the $\eta \pi\pi$ final states, one
may challenge either the baryonium assignment or the initial
assumption.

In retrospect, there is no strong experimental evidence that the
$p\bar p$ threshold enhancement and X(1835) are the same
resonance. Very probably they have completely different underlying
structures. In fact we find that X(1835) has a natural
interpretation as $\eta^\prime$'s second radial excitation. In
this short note, we shall discuss its mass, total width,
production rate and decay pattern to convince readers of this
assignment.

\section{Mass, Decay Width and Production Rate}

There are nine low-lying pseudoscalar mesons $\pi, K, \eta,
\eta'$. The mass splitting between $\eta$ and $\eta'$ is mainly
caused by the axial anomaly. In the large $N_c$ limit, the
contribution from the anomaly vanishes \cite{dynamics}. Then these
nine states would form a good nonet in the limit of exact SU(3)
flavor symmetry.

For the radial excitations of $\pi, K, \eta, \eta'$, the dominant
part of their masses comes from nonperturbative QCD interaction,
which is universal for them and much bigger than their mass
splitting caused by different current quark mass. With nodes in
their radial wave functions, one naively expects the axial anomaly
will not affect the mass of $\eta'$'s radial excitations
significantly. In other words, the radial excitations of $\pi, K,
\eta, \eta'$ mesons tend to form a good nonet. In fact, all
members of their first radial excitations are known to lie close
to each other from PDG \cite{pdg}. Their masses are $\pi(1300\pm
100), K(1460), \eta(1295), \eta'(1475)$. There may exist nearly
ideal mixing between the two bare isoscalar states. Such a mixing
enhances the $s\bar s$ component in $\eta'(1475)$ and causes the
proximity of the masses of $\pi(1300)$ and $\eta(1295)$
\cite{pdg}. Without mixing, $\eta'(1475)$ would easily decay into
$\eta'\pi\pi$ final states. After mixing, $\eta'(1475)$'s wave
function contains a large component of $s\bar s$. So its dominant
decay modes are $K\bar K\pi$.

For the second radial excitations, we have $\pi(1800), K(1830),
\eta(1760)$ \cite{pdg}. Only $\eta'$'s second radial excitation is
missing. If this missing state is observed around 1835 MeV, it
will not be a surprise. We suggest the recently observed resonance
X(1835) as $\eta'$'s second radial excitation. X(1835) can easily
decay into $\eta'\pi\pi$ as $\eta'$'s radial excitation while the
mode $\eta\pi\pi$ is disfavored. From PDG,
$\Gamma\left(\eta(1295)\right)=(55\pm 5)$ MeV,
$\Gamma\left(\eta'(1475)\right)=(50\sim 90)$ MeV,
$\Gamma\left(\eta(1760)\right)=(60\pm 16)$ MeV. If X(1835) is
$\eta'$'s second radial excitation, the measured width $\Gamma
\left(X(1835)\right)=(67.7\pm 20.3\pm 7.7)$ MeV is also very
natural.

$J/\psi$ decays into $\gamma\eta'$ more easily than into
$\gamma\eta$ because intermediate virtual gluons are
flavor-neutral and $\eta'$ meson is mainly a SU(3) flavor singlet.
From PDG \cite{pdg}, the branching ratio $B(J/\psi \to \gamma
\eta')=(4.31\pm 0.3)\times 10^{-3}$. Through $\eta_1$ and $\eta_8$
mixing, the branching ratio $B(J/\psi \to \gamma \eta)=(8.6\pm
0.8)\times 10^{-4}$, which is a factor of five smaller. The
radiative decay $J/\psi \to \gamma \eta(1295)$ has not been
reported yet. The branching ratio of $B(J/\psi \to \gamma
\left(\eta(1405)+\eta'(1475)\right))=(4.8\pm 0.8)\times 10^{-3}$,
which is very large. According to PDG, $\eta(1405)$ and
$\eta'(1475)$ are two different states. We simply take $B(J/\psi
\to \gamma \eta'(1475))=(2.4\pm 0.8)\times 10^{-3}$. It's
interesting to note that the radiative decay of $J/\psi$ into
$\eta'$'s first radial excitation is not suppressed severely.
Therefore, there is no reason to expect strong suppression of the
decay $J/\psi \to \gamma \eta'(1835)$.

Naively we assume a suppression factor of three compared with
$J/\psi \to \gamma \eta'(1475)$ and arrive at
\begin{equation}
B(J/\psi \to \gamma \eta'(1835))\sim 0.8 \times 10^{-3}\;.
\end{equation}
Experimentally BES measured the product branching fraction
\cite{bes1}:
\begin{equation}
B(J/\psi \to \gamma \eta'(1835)) B(\eta'(1835)\to \pi^+ \pi^-
\eta^{\prime}) = (2.2 \pm 0.4(stat) \pm 0.4(syst)) \times 10^{-4}
\;.
\end{equation}
If we further assume $B(\eta'(1835)\to \pi^+ \pi^- \eta^{\prime})
\sim 40\%$, we have
\begin{equation}
B(J/\psi \to \gamma \eta'(1835))\sim 0.6 \times 10^{-3}\;,
\end{equation}
which is quite consistent with our naive expectation. Through the
mixing of $\eta_8$'s and $\eta_1$'s bare second radial
excitations, $J/\psi$ can also decay into $\gamma \eta(1760)$.
Similar to the $J/\psi\to \gamma\eta$ case, we assume its
branching ratio of $J/\psi \to \gamma \eta(1760)$ is suppressed by
a factor five compared to $J/\psi\to \gamma\eta'(1835)$, which is
$\sim 1.2 \times 10^{-4}$. Experimentally this branching ratio is
measured to be
\begin{equation}
B(J/\psi \to \gamma \eta(1760))= (1.3\pm 0.9) \times 10^{-4}\;.
\end{equation}
In other words, the radiative branching ratio of $X(1835)$ is
consistent with its assignment as $\eta'$'s second radial
excitation.

\section{Effective Lagrangian for $X(1835)\to \eta'\pi\pi$ Decay Mode}

In this section we discuss S-wave decay mode $X(1835)\to
\eta'\pi\pi$. Based on SU(3) flavor symmetry, we can construct a
general effective Lagrangian.
\begin{eqnarray}\label{lag}
{\cal L}&=&g_1 \mbox{Tr} \left( P M\right) \mbox{Tr} \left(
M^2\right) + g_2 \mbox{Tr} \left( P M^2\right) \mbox{Tr} \left(
M\right) +g_3 \mbox{Tr} \left( P M^3\right) \nonumber\\&& +g_4
\mbox{Tr}\left( P \right)\mbox{Tr}\left(M^3 \right) +g_5
\mbox{Tr}\left(P \right) \mbox{Tr}\left(M^2 \right)
\mbox{Tr}\left( M \right)  \nonumber\\&& +g_6 \mbox{Tr}\left( P
\right) \mbox{Tr}\left( M \right) \mbox{Tr}\left(M \right)
\mbox{Tr}\left(M \right) +g_7 \mbox{Tr} \left( P M\right)
\mbox{Tr} \left( M\right) \mbox{Tr} \left( M\right)
\end{eqnarray}
where the matrix $M$ is the ground state pseudoscalar nonet and
$P$ is its radial excitation:
\begin{eqnarray}\label{m}
M=\left(
\begin{array}{ccc}
{\pi^0\over \sqrt{2}}+{\eta_8\over \sqrt{6}}+{\eta_1\over \sqrt{3}}&\pi^+ &K^+\\
\pi^-&-{\pi^0\over \sqrt{2}}+{\eta_8\over \sqrt{6}}+{\eta_1\over \sqrt{3}}&K^0\\
K^-&{\bar K}^0&-{2\over \sqrt{6}}\eta_8+{\eta_1\over \sqrt{3}}
\end{array}
\right),
\end{eqnarray}
\begin{eqnarray}\label{p}
P=\left(
\begin{array}{ccc}
{\pi^0(1800)\over \sqrt{2}}+{\eta(1760)\over
\sqrt{6}}+{\eta'(1835)\over \sqrt{3}}&\pi^+(1800)
&K^+(1830)\\
\pi^-(1800)&-{\pi^0(1800)\over \sqrt{2}}+{\eta(1760)\over \sqrt{6}}+{\eta'(1835)\over \sqrt{3}}
&K^0(1830)\\
K^-(1830)&{\bar K}^0(1830)&-{2\over
\sqrt{6}}\eta(1760)+{\eta'(1835)\over \sqrt{3}}
\end{array}
\right).
\end{eqnarray}
We have explicitly assumed X(1835) is $\eta'$'s second radial
excitation in Eq. (\ref{p}). In Eq. (\ref{m}), $\eta_{1,8}$
denotes SU(3) flavor octet and singlet member.

The $g_6$ and $g_7$ pieces in Eq. (\ref{lag}) involve two or three
$\eta_1$ mesons. Hence these modes are kinematically forbidden.
The pieces with $g_4$ and $g_5$ describe $\eta'(1835)$'s decay
only. With these terms only, the octet members of the second
radial excitations would not decay, in contradiction with
available experimental data. Hence their contribution should be
small. The $g_2$ term requires the decay final states of every
member within the $\eta'(1835)$ nonet contain the SU(3) flavor
singlet $\eta_1$, which is certainly not the case according to PDG
\cite{pdg}. Therefore, this term should not play a dominant role.
Now we are left with only two pieces.

If we keep the $g_3$ term only, we have
\begin{eqnarray}\label{22}
{\cal L}_{g_3}&=&{g_3\over 6} \eta'(1835)\cdot \{ \left(6\eta_1
+3\sqrt{2}\eta_8\right)\cdot \left( \pi^0\pi^0+2\pi^+\pi^-\right)
\nonumber \\&& +6\sqrt{3}\left( {\bar K}^0 K^+\pi^- +K^0 K^-
\pi^+\right) +3\sqrt{6}\pi^0 \left(K^+K^--K^0{\bar K}^0\right)
\nonumber \\&&
+\left(12\eta_1-3\sqrt{2}\eta_8\right)\left(K^+K^--K^0{\bar
K}^0\right) +2\eta_1^3 +6\eta_1\eta_8^2-\sqrt{2}\eta_8^3 \}
+\cdots
\end{eqnarray}
Naively one finds the coupling between $\eta'(1835)$ and
$\eta_1\pi\pi$ is a factor of $\sqrt{2}$ larger than that between
$\eta'(1835)$ and $\eta_8\pi\pi$. However the physical states are
$\eta, \eta^\prime$, which is a mixture of $\eta_1, \eta_8$:
\begin{eqnarray}\nonumber
|\eta\rangle=\cos\theta |\eta_8\rangle -\sin\theta |\eta_1\rangle
\;,\\
|\eta^\prime\rangle=\sin\theta |\eta_8\rangle +\cos\theta
|\eta_1\rangle
\end{eqnarray}
with the mixing angle $\theta\approx -{\pi\over 9}$ \cite{pdg}.
After inserting the above expressions into Eq. (\ref{22}) we have
\begin{equation}
{\cal L}_{g_3}\sim \eta'(1835)\cdot \left( {\pi^0\pi^0\over
2}+\pi^+\pi^-\right)\cdot \left(0.7 \eta^\prime +1.0 \eta \right).
\end{equation}
It's clear that (1) the decay width of $\eta'(1835)\to\eta^\prime
\pi^0\pi^0$ mode is half of $\eta^\prime \pi^+\pi^-$ decay width.
BES may be able to measure it; (2) the decay width of $\eta\pi\pi$
modes are a factor of two bigger than that of $\eta^\prime \pi\pi$
modes even if we ignore the larger phase space; (3) the branching
ratio of $\eta'(1835)\to {\bar K}^0 K^+\pi^- +K^0 K^- \pi^+$ is
nearly the same as that of $\eta'\pi^+\pi^-$. BES's
non-observation of $\eta \pi\pi$ and ${\bar K}^0 K^+\pi^- +K^0 K^-
\pi^+$ modes strongly indicate $g_3$ term does not play a dominant
role when $\eta'(1835)$ decays.

With the above argument, we conclude the $g_1$ piece in Eq.
(\ref{lag}) plays the dominant role when the $\eta'(1835)$ nonet
decays into three pseudoscalar mesons via S-wave. After expanding
this term we have
\begin{eqnarray}
{\cal L}_{g_1}= \left(\eta'(1835)\eta_1 + \eta(1760)\eta_8\right)
\left(2\pi^+\pi^-+\pi^0\pi^0+2K^+K^-+2K^0{\bar K}^0 +\eta^2
+{\eta^\prime}^2\right)+\cdots
\end{eqnarray}
From the above equation, the main decay modes of $\eta'(1835)$ is
$\eta'\pi^+\pi^-$ and $\eta'\pi^0\pi^0$. $\eta'K^+K^-$ and
$\eta'K^0{\bar K}^0$ modes are kinematically suppressed.

We would like to emphasize the decay mechanism from the $g_1$
piece is quite general for the decays of radial excitations. For
example, $\psi(2S)$ and $\Upsilon(2S)$ decay into $J/\psi\pi\pi$
and $\Upsilon\pi\pi$ in the same way.

\section{Discussion}

In short summary, we have noticed that there does not exist strong
experimental evidence that the $p\bar p$ threshold enhancement and
X(1835) have the same underlying structure. Very probably they are
two different states even if the enhancement arises from a
sub-threshold resonance. We point out that the mass, total decay
width, production rate and decay pattern of X(1835) are consistent
with its assignment as $\eta'$'s second radial excitation. Its
decay mode $X(1835)\to \eta'\pi^+\pi^-$ occurs through the
emission of a pair of S-wave pions, which is quite general for the
double-pion decays of ordinary radial excitations.
$\eta'\pi^0\pi^0$ mode should be within reach of BES detectors.
The confirmation of the absence of decay mode $X(1835)\to
\eta\pi\pi$ in the future experimental search by BES collaboration
will be a strong support of this classification. It is also very
interesting for BES to (1) search X(1835) in the $\eta' 4\pi$
modes; (2) look for the radiative decay $J/\psi \to \gamma
\eta(1295)$; (3) search $\eta'(1475)$ in the $\eta' \pi\pi$ final
states.

\section{Acknowledgments}

T.H. was supported by the National Natural Science Foundation of
China under Grants 10275070 and 10475084. S.L.Z. was supported by
the National Natural Science Foundation of China under Grants
10375003 and 10421003, Ministry of Education of China, FANEDD,
KJCX2-SW-N10, Key Grant Project of Chinese Ministry of Education
(NO 305001) and SRF for ROCS, SEM.


\end{document}